\documentclass[aps,prb,superscriptaddress,showpacs,twocolumn]{revtex4-1}        
\usepackage{dcolumn}
\usepackage{multirow}
\usepackage{graphicx}
\usepackage{subfigure}
\usepackage{epsf}
\usepackage{epstopdf}
\usepackage{color}
\begin{document}
\title{Impurity-vacancy complexes and ferromagnetism in doped sesquioxides}
\author{Paola Alippi}
\affiliation{CNR-ISM,
Via Salaria, Km 29.300, I-00016 Monterotondo Scalo, Rome, Italy} 
\author{Maura Cesaria}
\affiliation{ Dipartimento di Matematica e Fisica ``Ennio De Giorgi'', 
Universit\`a del Salento, Via Arnesano, I-73100 Lecce, Italy}
\author{Vincenzo Fiorentini}
\affiliation{Dipartimento di Fisica, Universit\`a di Cagliari, Cittadella Universitaria, Monserrato, I-09042 Cagliari, Italy}
\affiliation{CNR-IOM, UOS Cagliari, Cittadella Universitaria, Monserrato, I-09042 Cagliari, Italy} 
\date{\today}
\begin{abstract}
Based on hybrid density-functional calculations, we propose that ferromagnetism in the prototypical bixbyite sesquioxide In$_2$O$_3$ doped with Cr is due to Cr-oxygen vacancy complexes, while isolated Cr cannot support carrier-mediated magnetic coupling. Our proposal is consistent with  experimental facts such as the onset of ferromagnetism in O-lean conditions only, the low or vanishing net moment in unintentionally doped material, and its increase upon intentional doping. 
\end{abstract} 
\pacs{75.50.Pp,71.55.-i,71.15.Mb,61.72.Bb}
\maketitle

Transparent conducting oxides are key ingredients of various electronic technologies, and among them bixbyite sesquioxides X$_2$O$_3$ (with X=In, Y, rare earths, as well as Ga and Al in some variants) are under much current investigation. A potentially momentous addition to their functionalities is room-temperature ferromagnetism (FM), a prototypical example thereof being Cr-doped indium oxide (Cr:In$_{2}$O$_{3}$). \cite{PhilipNM06,biblio2,biblio3} The understanding of the FM mechanism, however, is still incomplete. In this paper we study  Cr:In$_{2}$O$_{3}$ using  hybrid density-functional theory, and find that the spin-polarized  states of isolated Cr are inaccessible to doping, hence incompatible with FM activated by extrinsic carriers. We then propose that FM be due instead to Cr-V$_{\rm O}$ complexes, whose properties match, among others, the strong experimental indications that Cr FM is only achieved in O-deficient conditions. Our conclusions hardly depend on the host cation, so we expect them to apply to all bixbyite sesquioxides. Since  FM and oxygen deficiency correlate experimentally  for most transition-metal dopants (e.g. Mn \cite{biblio3} and Ni \cite{peleckis}), the general concept may be relevant to transition-metal-doped sesquioxides in general.
 
Cr:In$_2$O$_3$ is FM with above-room-temperature Curie point \cite{PhilipNM06,biblio2,biblio3}  in O-lean  preparation conditions, with saturation magnetization increasing with free carrier concentration. In a diluted-magnetic-semiconductor-like \cite{DietlNM03} picture, it was suggested \cite{HuangEPL09,ZungerPRL09}  that  FM   is  due to extrinsic  (i.e. provided by doping or defects) carriers populating spin-polarized Cr-induced levels, resonant  with the  host conduction band. 
A key ingredient of a theoretical  discussion of magnetic coupling between Cr  impurities  is the accurate description  of the position of Cr-induced states relative to the bulk bands edges of In$_2$O$_3$, which of course should also be accurately described. This is a contentious point in ab initio density-functional theory (DFT) defects calculations. Semilocal functionals, e.g. the local  (LDA)  and  gradient-corrected  (GGA) approximations,  underestimate the fundamental gap and may misplace the extrinsic levels within the host band structure. Indeed,  non-local   empirical potentials (NLEP)  and ``+U'' corrections had to be added to GGA to obtain the current picture \cite{NLEP} of Cr:In$_{2}$O$_{3}$. To cure the  drawbacks of semilocal functionals, we use the Heyd-Scuseria-Ernzerhof (HSE) hybrid  exchange-correlation functional,\cite{HSE06} a GGA\cite{PBE}  functional modified to include  a fraction $\alpha_{\rm HF}$ of non-local short-range Hartree-Fock exact 
exchange screened by a model dielectric function. Indeed, HSE outperforms standard local and semilocal functionals in many instances of materials physics, including defect properties.\cite{stroppa,fegan}

We perform collinear spin-polarized total-energy, force, and band calculations for In$_2$O$_3$ in various states (undoped, Cr- and Sn-doped, O-deficient, and containing Cr-O vacancy complexes) with  the hybrid HSE  functional and  the projector augmented wave (PAW) method \cite{paw} as implemented in the VASP code.\cite{vasp}   We use an energy cutoff of 500 eV,  a $\Gamma$-centered 2$\times$2$\times$2 Monkhorst-Pack k-points grid, and treat In 4$d$ and Cr 3$d$ states as valence. All calculations are done at the experimental \cite{MarezioAC66}  lattice constant $a_{\rm eq}$=10.12 \AA.  
While the standard  HSE solid-state recipe is $\alpha_{\rm HF}$=0.25, here we examine the electronic structure for $\alpha_{\rm HF}$  in the range 0$\div$0.3,  with a standard screening wave-vector $\mu$=0.2 \AA$^{-1}$.  Full structural optimization is performed in all cases with a force threshold of  
0.02 eV/\AA. For $\alpha_{\rm HF}$=0.25
the Kohn-Sham band gap (valence band maximum (VBM) to conduction band minimum (CBM)) of bulk In$_2$O$_3$ is 2.9 eV, direct at $\Gamma$, as in other HSE calculations,\cite{KorberPRB10} 
 and close to  experiment.\cite{KingPRB09}
The valence band is  $\simeq$6  eV wide as in  experiment,\cite{ChristouJAP2000} and O 2$p$-like. The CBM is free electron-like and of mainly  In 5$s$ character. In 4$d$ states are around 13.5 eV below the VBM, also close to  experiment.\cite{KleinAPL09} The HSE electronic structure of  In$_2$O$_3$ is thus a good starting point to study Cr doping.

 In$_2$O$_3$ is a bixbyite,\cite{marsella}  with   six-fold coordinated cations  occupying the Wyckoff \cite{wyckoff} sites $8b$ and $24d$  (referred to below as $b$ and $d$). The $b$ cations have six O neighbors at the same distance, the $d$'s three pairs of O's at three distinct (but similar) distances. All O's are four-coordinated with four different bond lengths.
The isolated Cr impurity replaces one In atom at the $b$ or the $d$ site in a doubled cell (80 atoms), so the Cr density is 3.1\%. In both cases, the  Cr-O bonds are shorter ($\simeq$--7\%) than In-O bonds. The favored site, which we deal with in the following, is Cr$_b$, by 55 meV over Cr$_d$ (whose electronic structure is very similar, aside from small degeneracy lifting due to lower symmetry).

\begin{figure}[h]
\centering
\includegraphics[clip,width=8.5cm]{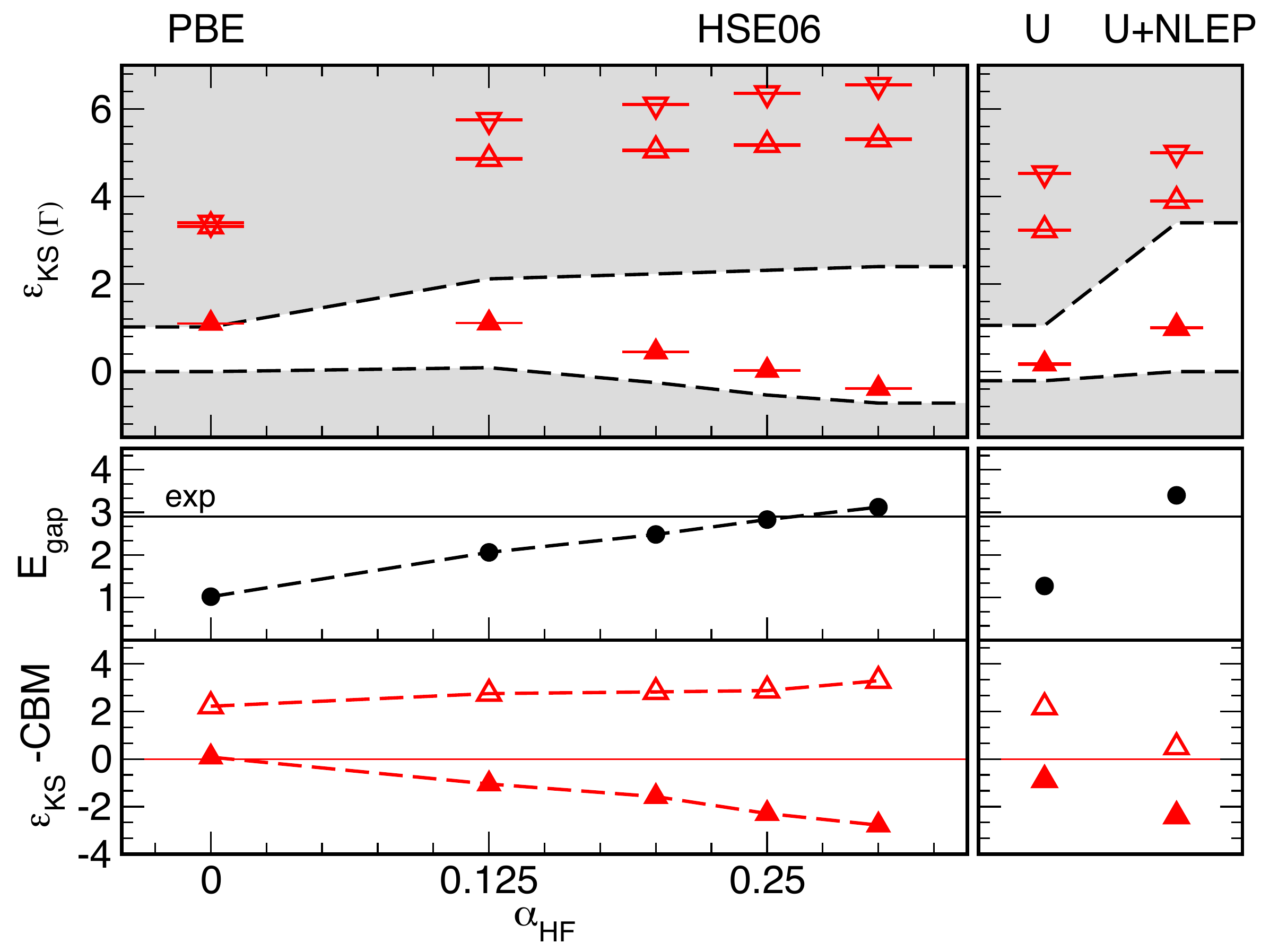}
\caption{(Color online) Left panel, levels as function of $\alpha_{\rm HF}$: Cr$_{b}$ levels within the In$_{2}$O$_{3}$ bands (top);  In$_{2}$O$_{3}$ band gap  (center); last occupied and first empty Cr$_{b}$ states (bottom) referred to CBM. Right panel: same quantities as left panel in  GGA+U (U$_{\rm In}$=4.5  eV, U$_{\rm Cr}$=2  eV) and  in GGA+U/NLEP (Ref.\onlinecite{ZungerPRL09}). 
 Legend: majority (minority) states are up (down) triangles, occupied (empty) states are filled (empty) triangles. Black circles are gap values. Energies in eV, at   $\Gamma$ point.}
\label{figAlpha}
\end{figure}

The $d^3$ configuration of trivalent Cr splits into $t_{2g}$ and $e_g$ states in the octahedral crystal field of In$_2$O$_3$. Its electronic structure as calculated in GGA, HSE, GGA+U, and NLEP \cite{ZungerPRL09} is depicted in Fig.\ref{figAlpha}
This is is crucial to  carrier-mediated FM,\cite{payne} in the light of earlier proposals\cite{ZungerPRL09} that Cr levels are close to the CBM and become occupied by extrinsic carriers. Indeed, the main result of Fig.\ref{figAlpha} is that within HSE calculations there is no Cr level in the vicinity of the CBM, roughly independently of $\alpha_{\rm HF}$, with the  empty majority  $e_g$ about 2.5 eV above the CBM, and   empty minority states all upward of 3 eV.  There is, in other words, no  near-edge donor state. The only Cr state in the gap is the occupied $d^3$ $t_{2g}$, which  nears  the VBM as $\alpha_{\rm HF}$ increases and might thus be viable as an acceptor; however, this is irrelevant in practice in In$_2$O$_3$, where $p$-doping is as difficult as $n$-doping is endemic. (In the top panels of Fig.\ref{figAlpha} each set of levels is referred to its own internal potential, except NLEP whereby the VBM is set to zero; clearly, however, only the differences between levels for {\it each individual} $\alpha_{\rm HF}$, or for GGA, GGA+U,  etc., are significant.)

As expected, the hybrid functional opens up the band gap nearly linearly  in
$\alpha_{\rm HF}$ (Fig.\ref{figAlpha}, central panel), matching  experiment at $\alpha_{\rm HF}$=0.25. As already mentioned, the  Cr empty states follow the blue shift of the CBM, while the occupied states follow the 
VBM (Fig.\ref{figAlpha}, top  panel). Occupied states dive away from 
the CBM already at low $\alpha_{\rm HF}$, as expected from self-interaction removal, whereas empty Cr states hardly move with respect to the  CBM. In particular, the lowest unoccupied majority  levels are about 2.5 eV from CBM  at all $\alpha_{\rm HF}$, so the exact value of $\alpha_{\rm HF}$ is not  critical. 

Empty and occupied  Cr-states separate drastically, from around 2 eV in GGA to 5 eV in HSE (Fig.\ref{figAlpha}, bottom panel). This separation is experimentally unknown in Cr:In$_{2}$O$_{3}$ but it was measured  from combined X-ray photoemission and Brehmsstrahlung spectroscopy \cite{cr2o3} to be  around 6 eV  for Cr$_2$O$_3$, where  Cr is also  trivalent and  octahedrally coordinated. We found that this value for Cr$_2$O$_3$ is reproduced by $\alpha_{\rm HF}$=0.25,  providing an independent countercheck of, and support  to this  recipe. 

For comparison, we studied Cr$_b$ by the GGA+U method (Fig.\ref{figAlpha}, right panel) with U values  as in the NLEP calculation.\cite{ZungerPRL09} GGA+U  yields a gap of 1.3 eV, less than half the experimental value. This is unsurprising as the relative orbital and spin polarization of the gap-edge states, respectively small and  non-existent, offer limited leverage to  U.) Occupied Cr-induced levels are at 0.5 eV above the VBM, and empty states $\simeq$2 eV above the CBM.  Overall, these  results are roughly similar to HSE with $\alpha_{\rm HF}$$\sim$0.1. Comparison with GGA+U/NLEP results\cite{ZungerPRL09} (also in Fig.\ref{figAlpha}, right panel) shows that the relative shift of CBM and dopant potential is due to the NLEP corrections, which keep the Cr-levels close to the CBM while the gap is opened up. (The NLEP gap was tuned \cite{ZungerPRL09} to the optical  onset of 3.5 eV, rather than to the lowest dipole-forbidden gap  of 2.9 eV \cite{KingPRB09}. HSE at $\alpha_{\rm HF}$=0.25, in turn, gives both gaps in  agreement  with experiment.)

The implications of the result for  carrier-mediated FM are obvious.  While Ref.\onlinecite{ZungerPRL09} reports empty minority Cr $e_{g}$ levels close in energy 
to the CBM and easily occupied  by free carriers (density 10$^{21}$ cm$^{-3}$), thus producing itinerant FM, our   results indicate  that carrier-mediated FM cannot result from such  occupation of Cr states in the presence of Cr alone. Empty Cr states are over 2.5 eV  above the CBM,  so that  their filling by electrons provided by  In-substitutional bond-saturating  shallow donors such as Sn is impossible, as the Fermi level cannot  rise so high  at any Sn concentrations. We verified by calculations that Sn, as expected of a shallow donor, hardly modifies the band structure and just induces a near-CBM shallow level (effectively  lowering the  CBM by 40 meV at  3.1\% Sn) which does not interact  with any  Cr levels. Thus, the free carriers do not interact with any  Cr levels, and we are led to conclude  that  FM cannot be sustained  in In$_2$O$_3$ by Cr alone at any doping level.

To make progress, we note that experiments  \cite{PhilipNM06,biblio2,biblio3} indicate unambiguously that FM occurs in Cr:In$_{2}$O$_{3}$ only for O-lean growth conditions (i.e. low oxygen pressure) or after high-vacuum annealing, whereby O loss upon cooling is endemic (in fact, this makes it easier to achieve FM in practice).  FM is however suppressed entirely (see especially Fig.3c of Ref.\onlinecite{PhilipNM06}) at large O pressures or after O-rich annealing.   It is thus natural to suppose that   defects related to oxygen deficiency be  involved in establishing FM (as e.g. for Co in ZnO \cite{cozno}). However, given that empty Cr states are energetically inaccessible, such defects should not be just charge providers, but rather interact with and modify the states of the Cr impurity. As a minimal sample, we study the isolated vacancy V$_{\rm O}$ and the  Cr$_{b}$-V$_{\rm O}$ complex  in their neutral state (the relevant doping is $n$-type, so vacancy deep donors never charge positively). We assume  concentrations  of 2\% and 3\% respectively for V$_{\rm O}$ and  Cr, i.e.  one vacancy and one Cr per simulation cell (density of 9$\times$10$^{20}$ cm$^{-3}$ each). 
 
\begin{figure}[ht]
\centering
\includegraphics[clip,width=8cm]{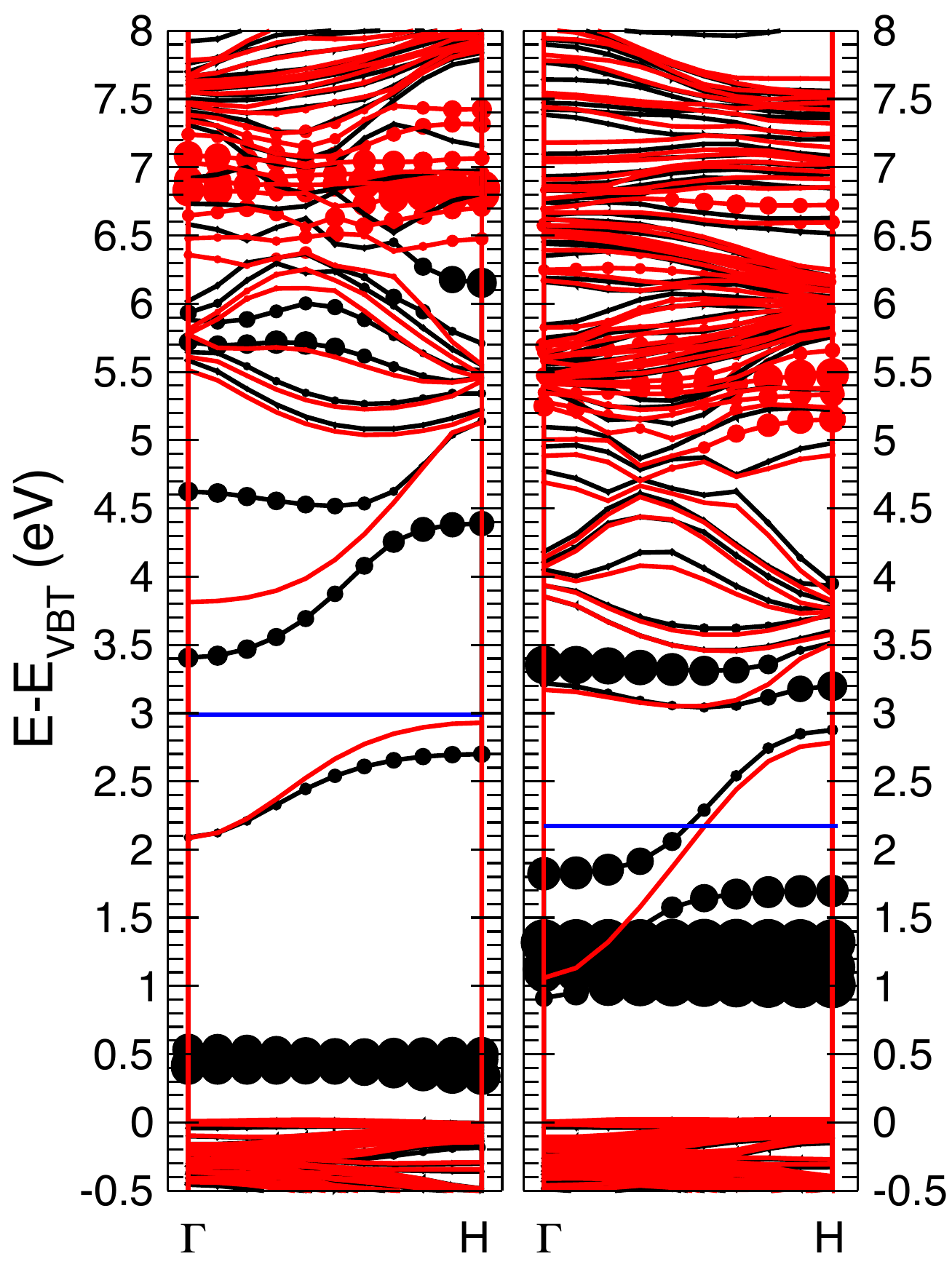}
\caption{(Color online) HSE (left) and GGA+U (right) band structure of the Cr$_{b}$-V$_{\rm O}$ complex. Majority states are black, minority are red (gray). Size of dots is proportional to projection on Cr. Energy zero in both panels is valence band edge. Horizontal line is the Fermi energy.}
\label{figcrvo}
\end{figure}


The  isolated vacancy  gives rise to a 
 moderately dispersed, mid-gap, fully occupied, non-spin-polarized
  state given by  the totally-symmetric combination of In dangling states.
On the other hand, as shown in Fig.\ref{figcrvo}, left panel, the Cr$_{b}$-V$_{\rm O}$ complex --i.e. the vacant site with one Cr$_{b}$ and three In neighbors-- has two spin-split  occupied gap states, one  of  Cr-In character in the majority channel (about 25\% Cr), and one purely In-like  in the minority.  The majority state is lower in energy across the  Brillouin zone, due to Hund's coupling with Cr $t_{2g}$ majority states. Due to on-site exchange  splitting, only Cr majority states  are  available in this energy range. Both the minority and majority gap states are filled, hence the  magnetic moment is 3 $\mu_{B}$ from the Cr $t_{2g}$ core. 

The  HSE electronic structure of the complex is conducive to long-range magnetic coupling, and compatible with experiment, as we will discuss momentarily.
Before that, we  note that in the GGA+U band structure in Fig.\ref{figcrvo}, right panel,  a heavily Cr-like polarized state drops in energy and becomes  occupied, while a mixed Cr-In doublet hosts the second electron and remains half-occupied. This stems (erroneously, by comparison with HSE) from the smaller gap, more dispersed vacancy states, and artificially stronger coupling of empty $e_{g}$ and occupied t$_{2g}$. The partially occupied polarized states imply doping-independent itinerant coupling, contrary to experiments. 

Coming back  to the HSE  electronic structure, the spectrum is gapped, and the occupied states cannot provide  itinerant coupling; thus, without direct doping and for V$_{\rm O}$ density near that of Cr, the system is paramagnetic and semiconducting. However, the Cr-induced  spin polarization propagates to  conduction band states (Fig.\ref{figcrvo}, left panel), which are now  significantly Cr-like, spin-polarized, and (by definition) within reach of intentional or accidental doping: charge added into these states will be delocalized and polarized, and will  produce an itinerant coupling between neighboring Cr centers. The nominally undoped material can be weakly FM with small magnetization when O-deficient, due to  unintentional doping or more likely to impurity bands due to V$_{\rm O}$ excess over Cr, which is  typical in experiment. Conduction will be hopping-like, as indeed often observed \cite{peleckis,mott1}.

Upon intentional $n$ doping, the experimental magnetization increases significantly (up to 1.5 $\mu_{B}$ or so), but  always falls short of the saturation value.  In our picture, this increase should occur as the electron density is raised up to 1.5$\div$2$\times$10$^{21}$ cm$^{-3}$, whereby only the majority-Cr-like CBM state is occupied (as can be estimated using a free-electron--like density of states and assuming rigid-band behavior). This matches roughly the typical  2:1 ratio of free-electron to Cr concentration in FM  samples.\cite{PhilipNM06} Above that density, the minority state will progressively  fill up and compensate the majority. While it  proved  computationally impossible to calculate directly the magnetic coupling and  critical temperature at this Cr density, it is virtually guaranteed that the itinerant coupling will be FM, and not antiferromagnetic, given the plethora of available states for FM hopping in the standard virtual-superexchange picture.
We note in closing that our picture  resembles in part that based  on carrier-mediated Cr coupling.\cite{ZungerPRL09} The key difference is that the  Cr-V$_{\rm O}$ complex must be present to bring $e_{g}$-like states within reach of doping. Similarly, the correlation of FM and oxygen deficiency has been previously attributed to oxygen vacancies as providers of mobile charge; here, instead, the main role of vacancies is that of co-forming the magnetic building block whose polarized states can be occupied by extrinsic charge, provided by dopants or vacancies in excess over the dopant concentration.

We observe that  Cr$_{b}$ and V$_{\rm O}$ which are not first neighbors do not interact (their  electronic structure is essentially the superposition of those of the isolated defects). Cr-V$_{\rm O}$  is thus  the effective magnetic-dopant unit in this system, and should be thermodynamically  bound to act as such. Its binding energy
$B$=$|E_{\rm CrV_{\rm O}}+E_{\rm bulk}|-
|E_{\rm Cr}+E_{\rm V_0}|$ should be positive
 (the $E$'s are the energies of supercells containing the Cr-V$_{\rm O}$ complex,  no defect, and each of the two isolated defects).
At the  In$_{2}$O$_{3}$ equilibrium volume and equal Cr and V$_{\rm O}$ densities, $B$$\simeq$0 to numerical accuracy.
Configurational entropy decreases upon complex formation (by of order 1 k$_{\rm B}$/complex),  weakly favoring isolated defects. Vibrational entropy should increase on account of the softer modes of Cr in the complex (three bonds to O) than as substitutional (four bonds), but no  huge change is expected (in fact, no change at all in a model assuming only In$\rightarrow$Cr mass changes at  fixed force constants). It is then plausible to assume  that entropic contributions  cancel out to within  1 or 2 k$_{\rm B}$ either way, thus causing  concentration changes in the $\pm$5-10\% range at most. In this case, we end up with roughly equal populations of Cr-V complexes  (magnetically active) and isolated Cr (magnetically inactive). This suffices to cause itinerant magnetism, and in fact magnetizations at most of order a half the maximum of 3 $\mu_B$/Cr, as is generally observed upon doping.  

How interactions of Cr with a very large vacancy density will play out in detail (e.g. via  the local structure and bonding of multivacancy-Cr complexes, or impurity band formation, or entropic effects) cannot be assessed directly, as a configurational sampling of vacancy-Cr configurations is computationally prohibitive (for instance, we considered volume reduction as would occur at high oxygen deficiency  via the negative vacancy formation volume, but that does not change $B$ appreciably). Whatever the details,  within the present picture it remains true that O-rich growth conditions or oxidizing post-processing will suppress the magnetic coupling, as consistently observed,\cite{PhilipNM06,biblio2} by annihilating vacancies and hence Cr-V$_{\rm O}$'s.  As mentioned, normal handling of In$_{2}$O$_{3}$, such as cooling in air,  is effectively reducing, so O deficiency and hence magnetism will generally be easily realized. 

In summary,  the electronic structure of isolated Cr in In$_{2}$O$_{3}$ as provided by hybrid density functionals is not compatible with ferromagnetism. We proposed an alternative mechanism enabled by  Cr-V$_{\rm O}$ complexes and boosted by intentional  doping, which is consistent with  experimental results such as the low net moment in unintentionally doped material, FM onset only in O-lean conditions, moment increase upon intentional doping, and smaller than expected saturation moment. Our results should largely apply to sesquioxides sharing the bixbyite structure. Given the similar  FM-oxygen deficiency relation for other transition metal dopants, the general concept might be relevant to transition-metal-doped sesquioxides in general.

Work supported in part by MIUR-PRIN 2010 project {\it Oxide}, Fondazione Banco di Sardegna and CINECA-ISCRA grants.
MC thanks her PhD supervisors M. Martino and A. P. Caricato for their support and assistance. This paper is based in part on ideas, portions, and perspectives presented in MC's PhD thesis (University of Salento, Lecce, Italy (2012)).

\end{document}